# Evaluating the Performance of IPTV over Fixed WiMAX

Jamil Hamodi
Information Technology
Department
Shri Guru Gobind Singhji
Institute of Engineering and
Technology (SGGS)
Nanded, India

Khaled Salah
Computer Engineering
Department
Khalifa University of Science,
Technology and Research
(KUSTAR)
Sharjah, UAE

Ravindra Thool
Information Technology
Department
Shri Guru Gobind Singhji
Institute of Engineering and
Technology (SGGS)
Nanded, India

## ABSTRACT

IEEE specifies different modulation techniques for WiMAX; namely, BPSK, QPSK, 16 QAM and 64 QAM. This paper studies the performance of Internet Protocol Television (IPTV) over Fixed WiMAX system considering different combinations of digital modulation. The performance is studied taking into account a number of key system parameters which include the variation in the video coding, path-loss, scheduling service classes different rated codes in FEC channel coding. The performance study was conducted using OPNET simulation. The performance is studied in terms of packet lost, packet jitter delay, end-to-end delay, and network throughput. Simulation results show that higher order modulation and coding schemes (namely, 16 QAM and 64 QAM) yield better performance than that of QPSK.

## 1. INTRODUCTION

Worldwide Interoperability for Microwave Access (WiMAX) technology is one of the solutions of fourth-generation (4G) wireless network which provides high data rates for IP networks that is capable of offering high Quality of Service (QoS). The high data rate and Quality of Service (QoS) assurance provided by this standard has made it commercially viable to support multimedia applications such as video telephony, video gaming, and mobile TV broadcasting. WiMAX base station (BS) can provide broadband wireless access in a range up to 30 miles (50 km) for fixed stations and 3 to 10 miles (5 to 15 km) for mobile stations with a maximum data rate of up to 70 Mbps [1- 3].

The WiMAX standard product is specifically for fixed and Nomadic services. It was reviewed to address full mobility applications. Hence, Mobile WiMAX supports full mobility for nomadic and fixed systems. It addresses the following features [4]: offers high data rates; supports fixed, nomadic and mobile applications thereby converging the Fixed and mobile networks; and has flexible network architectures; in addition to its cost-effective and easy to deploy. Moreover, it can support point to point and point to multipoint connection also support IP based architecture; and has optimized handover which support full mobility application such as Voice over Internet Protocol (VoIP). It has also the power saving mechanism which increases the battery life of handheld devices.

Internet Protocol Television (IPTV) provides digital television services over IP for residential and business users at a lower cost. Moreover, IPTV is a system capable of receiving and displaying a video stream using Internet Protocol [5]. Users can get IPTV services anywhere and anytime to mobile devices. IPTV services can be classified by their type of content and services to [6]: **Video-on-Demand content:** In this IPTV service a customer is allowed to browse an online movie catalogue, to watch trailers, and to select a movie of interest. Unlike the case of live video, a customer can request or stop the video content at any time and is not bound by a particular TV schedule. The playout of the selected movie starts nearly instantaneously on the customer's TV or PC. **Live content:** In this IPTV service a customer is required to access a particular channel for the content at a specific time, similar to accessing a conventional TV channel. A customer cannot request to watch the content from the beginning if he or she joins the channel late. Similar to a live satellite broadcast, live content over IPTV can be a showing of a live event or a show encoded in real-time from a remote location, such as a soccer game. **Managed services:** It enables video content to be offered by the phone companies who operate the IPTV business or obtained from syndicated content providers, in which the content is usually well-managed in terms of the coding and playout quality, as well as in the selection of video titles. Bandwidth for delivery and customer equipment are arranged carefully for serving the best playout performance and quality to the customers. **Unmanaged services:** In this service the technology of IPTV itself enables playout of any live or on demand video content from any third party over the Internet. Therefore, nothing stops a customer from accessing video content directly from any third party online such as YouTube (or Google Video), individuals, or an organization. With a wide range of choices for content selection, obviously the unmanaged services have an advantage at the expense of non-guaranteed playout quality and performance.

In wireless communication systems, there are a number of factors affect the quality of a signal received by a user equipment. These factors namely the distance between the desired user and interfering base stations, path loss exponent, log-normal shadowing, short term Rayleigh fading and noise. In order to improve system capacity, peak data rate and coverage reliability, the signal transmitted to and by a particular user is modified to account for the signal quality variation through a process commonly referred to as link adaptation. Adaptive Modulation and Coding (AMC) has become a standard approach in recent developing wireless standards, including WiMAX. However, the idea behind AMC is to dynamically adapt the modulation and coding scheme to the channel conditions so as to achieve the highest spectral efficiency at all times [7].

Modulation coding in OFDMA can be chosen differently for each sub-carrier, and it can also change with time. In reality, in the IEEE802.16 standard, coherent modulation schemes are used starting from low efficiency modulations (BPSK 1/2) to very high efficiency (64-QAM 3/4) depending on the Signal-to-Noise Ratio (SINR). The amount of data transferred through a single channel depends on the variation in the modulation and coding scheme, this also leads to use the best modulation with lower data dropped and higher throughput. With the large number of consumers of QoS-enabled high rate services, it is required to have knowledge of performance parameters of IPTV (VoD) over Fixed WiMAX networks under the fixed type of modulation and coding to select the best combination.

This work is geared towards investigating the performance study of IPTV (VoD) over Fixed WiMAX networks when considering different modulation and coding schemes using simulation software OPNET Modeler. This work also is extending of the work in [8, 9] by include generation and integration of a streaming audio component, also provides a comparative study of performance of IPTV (VoD) over Fixed WiMAX under varying video coding and using different path-loss models and classes services under fixed types of modulation and coding in order to investigate and analyze the behaviour and performance of these models. OPNET provides comprehensive development of network models including all the necessary parameters that need to be reflected in the design procedure of PHY and/or MAC layers. A series of simulation scenarios under OPNET for broadband wireless communication is developed. The research work and results presented in this paper focus mainly on the use of real-time audio/video movies coded by different video coding (MPEG-4, AVC, and SVC) for modelling and using simulation IPTV deployment over Fixed WiMAX. This paper mainly aims to establish a comparative study of performance of IPTV (VoD) over Fixed WiMAX under varying video coding and using different path-loss models and classes' services under fixed types of modulation techniques in order to investigate and analyze the behaviour and performance of these models.

The remainder of this paper is organized as follows. Section 2 provides a brief state of the art, concluding with the architecture of IPTV over WiMAX networks. Section 3 gives relevant background and preliminaries on WiMAX modulation and coding schemes. Video traffic characteristics and requirements are highlighted in Section 4. Section 5 describes the practical steps to be taken prior to simulation. Simulation results and analysis obtained are provided in section 6. Section 7 is naturally the conclusion of the findings as a whole and a summation of this modest research endeavour.

## 2. LITERATURE REVIEW

Recently, there have been some works based on performance studies of video streaming over WiMAX networks. Many of research workers have explored WiMAX in the context of real-time and stored video applications. For example, Pandey et al. [5] Developed a model to dimension the network for IPTV service providers that offer VoD services to their customers in heterogeneous environments. The proposed modelling and simulation technique allows us to determine the optimum deployment conditions for a given number of potential IPTV users while satisfying predefined QoE measures. On other hand, Shehu et al. [10] Discussed issues regarding challenges for delivering IPTV over WiMAX. These issues include the challenges of QoS requirements. Also, they describe the transmission of IPTV services over WiMAX technology, and the impact of different parameters in WiMAX network when deploying this service. An intelligent controller has been designed based on fuzzy logic to analyze QoS requirements for delivering IPTV over WiMAX in [11] is used to analyze three parameters: jitter, losses and delays that affect the QoS for delivering IPTV services. The aim is to define a maximum value of link utilization among links of the network.

Hrudey et al. [12] used OPNET Simulation to design, characterize, and compare the performance of video streaming to WiMAX and ADSL. The simulation results indicate that ADSL exhibits behaviour approached the ideal values for the performance metrics while WiMAX demonstrates promising behaviour within the bounds of the defined metrics. The work in [13] is extending the work in [12] to include generation and integration of a streaming audio component, also enhances the protocol stack to include the real time protocol (RTP) layer. Network topology is redesigned to incorporate WiMAX mobility. Also, include characterization of WiMAX media access control (MAC) and physical (PHY) layer. Simulation scenarios are used to observe the impact on the four performance metrics. Gill et al. [14] used OPNET Simulation to compare the performance metrics between ADSL and WiMAX by varying the attributes of network objects such as traffic load and by customizing the physical characteristics to vary BLER, packet loss, delay, jitter, and throughput. Simulation results demonstrate considerable packet loss. ADSL exhibits considerably better performance than the WiMAX client stations.

Hamodi et al. [8] Used OPNET Simulation to design, characterize, and deployment the performance of video streaming over WiMAX under a different video codec (SVC, and AVC). The simulation results indicate that SVC video codec is an appropriate video codec for video streaming over WiMAX. The work in [9] is extending the work in [38] to investigate the performance of video streaming over WiMAX under two different terrain environments, namely Free Space, Outdoor to Indoor and Pedestrian. The simulation results indicate that, free space path loss model is a basic path loss model with all other parameters related to terrain and building density set as constant.

However, many of recent works explore the performance studies of WiMAX under different modulation and coding schemes. For example, Telagarapu et al. [15] analyzed the physical layer of WiMAX with different modulation techniques like BPSK, QPSK, QAM and comparison of QPSK modulation with and without Forward Error Correction methods. Islam et al. [4] Evaluated WiMAX system under different combinations of digital modulation (BPSK, QPSK, 64-QAM and 16-QAM) and different communication channels AWGN and fading channels (Rayleigh and Rician), and the WiMAX system incorporates Reed-Solomon (RS) encoder with Convolution encoder with ½ and 2/3 rated codes in FEC channel coding.

Bhunia et al. [16] presented an in-depth performance evaluation of mobile WiMAX is carried out using adaptive modulation and coding under the real-like simulation environment of OPNET. They have evaluated the performance parameters of mobile WiMAX with respect to different modulation and coding schemes. Their performance has been evaluated in terms of average throughput, average data-dropped, the MOS value of voice application and the BW usage in terms of UL data burst usage when deployed VoIP on WiMAX Networks. It has been observed that using lower order modulation and coding schemes, the system provides better performance in terms of throughput, data dropped and MOS at the cost of higher BW usage.

The architecture of IPTV over WiMAX networks described in this paper is based on the architecture introduced in Lloret et al. [17] as shown in Figure 1. The IPTV architecture over WiMAX network consists of five main subsystems. Head network is the first subsystem of the model. In this subsystem the servers store video content of any type of movies and audio content source. Different video types store the video source as national TV broadcasters, local broadcasters, Internet TV operations and any other future video broadcast service. The video content is delivered to the WiMAX network through the long distance, high capacity content distribution core network. The core network distributes the video flows from the header to the distribution network of the service provider. The distribution network which is the area contains the Base Station (BS). It goes from the end of the core network to the access network. The access network lets the user connect to the service provider and allows access to the multimedia content. The first requirement of an access network is to have enough bandwidth to support multiple IPTV channels for each subscriber while it allows other services (telephony and data). Finally, the customer network (Set-up-box) enables communication and information exchange between the computers and devices connected to the services offered by a service provider.

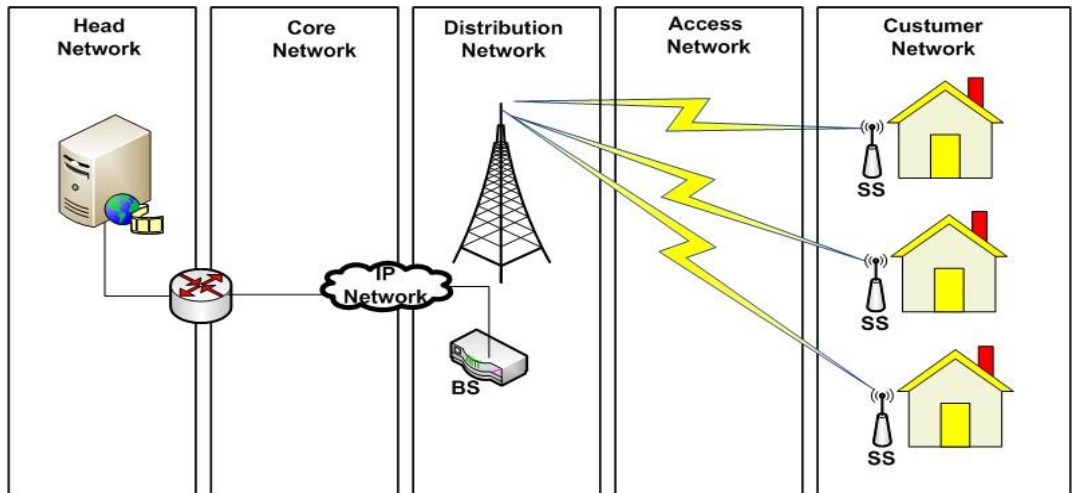

**Figure 1: System Model of IPTV over WiMAX Architecture**

## 3. BACKGROUND

This section provides relevant background and preliminaries pertaining to QoS support and the physical layer on WiMAX modulation and coding schemes. A variety of modulation and coding schemes have been supported in WiMAX, these modulation and coding allows the scheme to change on a burst-by-burst basis per link that is depending on channel conditions. Unlike uplink, the base station can estimate the channel quality based on the received signal quality. The channel quality feedback indicator helps the mobile provide the base station with feedback on the downlink channel quality. The base station scheduler can take into account the channel quality of each user's uplink and downlink and assign a modulation and coding scheme that maximizes the throughput for the available signal-to-noise ratio (SINR). Adaptive modulation and coding significantly increases the overall system capacity as it allows real-time trade-off between throughput and robustness on each link. WiMAX OFDM features multiple subcarriers ranging from a minimum of 256 up to 2048, each modulated with QPSK, 16 QAM, or

64 QAM modulation, where 64 QAM is optional on the uplink channel. The advantage of orthogonality is that it minimizes self-interference, a major source of error in receiving signals in wireless communications.

Channel coding schemes are used to help reduce the SINR requirements by recovering corrupted packets that may have been lost due to burst errors or frequency selecting fading. These schemes generate redundant bits to accompany information bits when transmitting over a channel. Coding schemes include convolution coding (CC) at various coding rates (1/2, 2/3 and 3/4) as well as conventional turbo codes (CTC) as various coding rates (1/2, 2/3, 3/4, and 5/6). The coding rate is the ratio of the encoded block size to the coded block size. The available coding rates for a given modulation scheme with the minimum signal to noise rate and the peak UL and DL data rates for 5 MHz channel mobile WiMAX with different information bits/symbol are listed in Table 1 [18, 19].

**Table 1. Mobile WiMAX PHY data rate and SINR for 5 MHz channel**

| Modulation Scheme | Coding | Information bits/symbol | Minimum SINR (dB) | Down-link rate (Mbps) | Uplink rate (Mbps) |
|---|---|---|---|---|---|
| QPSK | ½ | 1 | 5 | 3.17 | 2.28 |
| | ¾ | 1.5 | 8 | 4.75 | 3.43 |
| 16 QAM | ½ | 2 | 10.5 | 6.34 | 4.57 |
| | ¾ | 3 | 14 | 9.5 | 6.85 |
| 64 QAM | ½ | 3 | 16 | 9.5 | 6.85 |
| | 2/3 | 4 | 18 | 12.6 | 9.14 |
| | ¾ | 4 | 20 | 14.26 | 10.28 |

Additionally, WiMAX supports different signal bandwidths ranging from 1.25 to 20 MHz to facilitate transmission over longer ranges in different multipath environments. In wireless communication systems, information is transmitted between the transmitter and the receiver antenna by electromagnetic waves. During propagation, electromagnetic waves interact with the environment, thereby causing a reduction of signal strength. Another limiting factor for higher sustained throughput in wireless communications, especially when the terminal nodes have the mobility it is caused by reflections between a transmitter and receiver, viz, a propagation path between the transmitter and the receiver is regarded. The propagation path between the transmitter and the receiver may vary from simple line-of-sight (LOS) to very complex one due to diffraction, reflecting and scattering [20]. To estimate the performance of IPTV over WiMAX channels, propagation models in [16] are often used.

Path loss is an unwanted introduction of energy tending to interfere with the proper reception and reproduction of the signals during transfer from transmitter to receiver. This environment between the transmitter and the receiver in a wireless communication system has an important effect on the performance and to maintain QoS of the system. The path loss is an important element which must be kept within a predefined range in order to get the expected performance of the system. In addition, path loss models describe the signal attenuation between a transmitting and a receiving antenna as a function of the propagation distance and other parameters which provide details of the terrain profile required to estimate the attenuating signals. Path loss models represent a set of mathematical equations and algorithms which apply to radio signal propagation prediction in certain environments [21]. Path-loss is highly dependent on the propagation model, the common propagation models namely Free Space, Suburban Fixed (Erceg), Outdoor to Indoor and Pedestrian Environment and Vehicular Environment are given in Table 2. These models are used in Fixed WiMAX performance evaluation through OPNET simulation.

**Table 2. Path-loss models**

| Propagation Model | Mathematical formulation | Description |
|---|---|---|
| Free Space model | $P_{rx}(r) = P_{tx}G_{tx}G_{rx}/((4\pi)^2 r^2 L)$, $P_{rx}$ and $P_{tx}$ are received power in watts, respectively; $G_{rx}$ and $G_{tx}$ are the gain of the receiving and transmitting antennas, respectively; $L$ is the system-loss factor. | It is a mathematical model hardly applicable without considering the fading effect due to multi-path propagation. |
| Erceg's suburban fixed model | $PL = H + 10\gamma log_{10}\left(d/d_0\right) + X_f + X_h + s$, $PL$ is the instantaneous attenuation in dB, $H$ is the intercept and is given by free space path-loss at the desired frequency over a distance of $d_0$= 100m. $\gamma$ is a Gaussian random variable over the population of macro cells within each terrain category. $X_f$ and $X_h$ are the correlation factors of the model for the operating frequency and for the MS antenna height, respectively | It is based on extensive experimental data collected at 1.9 GHz in 95 macro cells of suburban areas across the United States. Very large cell size, base stations with high transmission power and higher antenna height. Subscriber stations are of very low mobility |
| Outdoor-to-indoor and pedestrian path-loss environment | $PL = 40log_{10}R + 30log_{10}f + 49$. $PL$ is the instantaneous attenuation in dB, $R$ is the distance between the base station and the mobile station in kilometers and $f$ is the carrier frequency | Small cell size, base stations with low antenna heights and low transmission power are located outdoors while pedestrian users are located on streets and inside buildings and residences |
| Vehicular environment | $PL = 40(1 - 4*10^{-3}*\Delta h_b)log_{10}R - 18log_{10}\Delta h_b - 21log_{10}f + 80dB.R$ is the distance between the base station and the mobile station, $f$ is the carrier frequency and $\Delta h_b$ is the base station antenna height in meters | Larger cells and higher transmitter power. All subscriber stations have a high mobility |

In general, IEEE 802.16 Medium Access Control (MAC) defines up to five separate service classes to provide QoS for various types of applications. The service classes include: Unsolicited Grant Scheme (UGS), Extended Real Time Polling Service (ertPS), Real Time Polling Service (rtPS),

Non Real Time Polling Service (nrtPS) and Best Effort Service (BE). Each service class has its own QoS parameters such as the way to request bandwidth, minimum throughput requirement and delay/jitter constraints. These service classes are [22- 24] described in Table3.

**Table 3. QoS parameters for each scheduling service**

|  | UGS | rtPS | ertPS | nrtPS | BE |
|---|---|---|---|---|---|
| Maximum Sustained Traffic Rate | X | X | X | X | X |
| Minimum Reserved Traffic Rate | -- | X | X | X | -- |
| Maximum Latency | X | X | X | -- | -- |
| Tolerated Jitter | X | -- | -- | -- | -- |
| Traffic Priority | -- | -- | -- | X | X |
| Application support | VoIP | Streaming audio and video | Voice with activity detection (VoIP) | File transfer protocol (FTP) | Data transfer, web browsing, etc. |

# 4. VIDEO TRAFIC CHARACTERISTICS AND REQUIREMENTS

This section discusses key issues related to performance metrics of video transmission and equipment needed for the deployment of IPTV. Performance metrics can be classified as objective and subjective quality measures. The aim of both methods is to obtain the video quality metrics. Objective measures that observe packet transmissions include packet loss, packet delay, packet jitter, and traffic load throughput rates. Other objective metrics that attempt to quantify video quality perceptions include the ITU video quality metric (VQM) and peak signal to noise ratio (PSNR) which measures the codec's quality of reconstruction. Establishing the performance of TV systems by using measurements that more directly anticipate the user perceptions, these measurements are the subjective video quality methods. It is concerned with how video is perceived by a viewer and designates his or her opinion on a particular video sequence. To evaluate those perceptions, a group of people watch the video and give it a quality score. The main idea of measuring subjective video quality is the same as in the Mean Opinion Score (MOS).

## 4.1 Quality of Service (QoS)

Quality of Service (QoS) requirement is very important for deploying IPTV and VoD as real time services over WiMAX networks. In order to assess the performance of video transmission systems, a suite of relevant performance metrics was identified to appropriately benchmark the system. Video on demand (VoD) deployments over WiMAX is affected by time varying bandwidth, packet delays, and losses. Since users expect high service quality regardless of the underlying network infrastructure, a number of metrics were collectively used to measure the video content streaming performance to ensure compliance and a user's quality of experience (QoE) [13].

We consider the following performance measures which are widely studied: packet loss ratio (PLR), packet delay (PD), packet jitter, and minimum throughput. The performance parameters affecting video have been shown in Table 4 as in [25].

**Table 4. Performance parameters for deploying VoD**

| Metrics | Mathematical formulation | Description | Acceptable |
|---|---|---|---|
| **Packet loss ratio (PLR)** | $$PLR = \left( \frac{lost_{packet}}{lost_{packet} - received_{packet}} \right)$$ | **PLR** is the corrupted, lost, or excessively delayed packets divided by the total number of packets expected at the video client station. | $10^{-3}$ |
| **Packet End-to-End Delay (E2E) (ms)** | $D_{E2E}= Q\ (d_{proc} + d_{queue} + d_{trans} + d_{prop})$ , where: $Q$ is the number of network elements between the media server and mobile station . $d_{proc}$ is the processing delay at a given network element. $d_{queue}$ is the queuing delay at a given network element. $d_{trans}$ is the transmission time of a packet on a given communication link between two network elements. $d_{prop}$ is the propagation delay across a given network link | **Packet delay** is the average packet transit time between the media server and the video client station. | <400 |
| **Packet delay variation (PDV) or Packet jitter (ms)** | $j_{pkt} = t_{actual} - t_{expected}$ , where: $t_{actual}$ is the actual packet reception time. $t_{expected}$ is the expected packet reception time. | **Packet jitter** is defined as the variability in packet delay within a given media at video client station. | <50 |
| **Throughput (bps)** | The **throughput** for variable bit rate (VBR) traffic loading is dynamic in nature and it is a function of the scene complexity and associated audio content. Variable bit rate (VBR) traffic loads is typically quoted as peak throughput ranges. | **Throughput** is defined as the traffic load that the media stream will add to the network. It can be measured in bits/sec | 221-5311 |

## 4.2 Video Equipments

The primary equipments needed to deploy IPTV (VoD) services over any IP networks are set-top boxes (STB) and a head-end server. The head-end server or known as VoD server is the source for all video content. The main functionality of set-top box (STB) is to unscramble the signal and present it on the TV [26].

## 5. SIMULATION MODEL

This section describes the simulation model used for analyzing the effect of Video on Demand (VoD) over the Fixed WiMAX Networks. The simulation was performed to evaluate the performance study of VoD over the Fixed WiMAX networks under different parameters: video codecs, path-loss models, and class's services under fixed types of modulation and coding techniques in order to investigate and analyze the behavior and performance of these models. Initially, topology shown in Figure 2 was considering. This topology has a server with a video encoder capable of transmitting video to the subscriber station (SS). It is also assumed that there are n WiMAX cells (BS) connected to the wired networks. An SS of each cell connects to the server and request the video stream in real-time. It is assumed that each SS at different distances from each BS so that each BS assigns different modulation and coding for SS. For example: QPSK ½ assign to SS in BS1, 16 QAM ¾ assign to SS in BS 2, and 64 QAM 2/3 assign to SS in BS n.

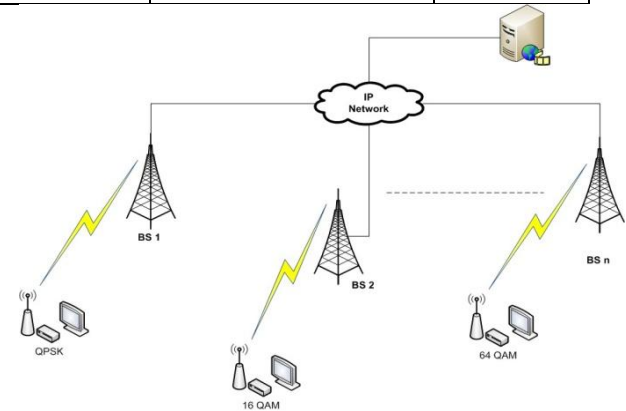

**Figure 2: Topology of IPTV (VOD) over WiMAX**

For the simulation, the popular MIL3 OPNET Modeler simulation is used [27]. Here the OPNET Modeler is used to facilitate the utilization of in-built models of commercially available network elements with reasonably accurate emulation of various real life network topologies.

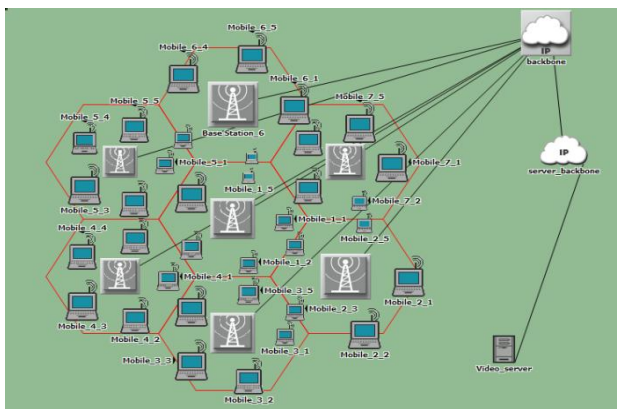

**Figure 3: OPNET Model of IPTV over Fixed WiMAX**

The network topology of a test - bed network was used in the simulation is exhibited in Figure 3. The simulation model of this case study network is deployed with 7-Hexagonal celled WiMAX with multiple subscriber stations in the range of a base station. The base stations are connected to the core network by an IP backbone. The IP backbone is connected to the video server via server backbone. The IP backbone and server backbone together represent the service provider company network. Single node SS was used for each cell is (mobile x_1). This node in each cell has been assigned to different modulation and coding scheme MCS depending on its distance from BS. For example, mobile 1_1 has QPSK ½ coding and etc. The common attributes used for network configuration are highlighted in Table 5.

**Table 5: Network Configuration Details**

| Network | Fixed WiMAX Network |
|---|---|
| Cell Radius | 0.2 Km |
| No. of Base Stations | 7 |
| No. of Subscriber Stations | 5 |
| IP Backbone Model | IP32_cloud |
| Video Server Model | PPP_sever |
| Link Model (BS-Backbone) | PPP_DS3 |
| Link Model (Backbone-server Backbone) | PPP_SONET_OC12 |
| Physical Layer Model | OFDM 5 MHz |
| Traffic Type of Services | Streaming Video |
| Application | Real Video streaming |
| Scheduling | rtPS |

Video streaming over wireless networks is a challenging task. This is due to the high bandwidth required and the delay sensitive nature of video more than most other types of application. Variable bit rate (VBR) video traffic models has emerged as an attractive alternative to overcome the drawbacks of CBR, which is accurately present the traffic characteristics and statistical properties of real video. Such as, it is costly, inefficient, and large delays [28]. As a result, a VBR video traces of 74 minutes Tokyo Olympics movie encoded by different codec: MPEG-4 part 2, H.264/AVC, and Scalable Video Coding (SVC) is used in simulation. This movie traces with different coding obtained from Arizona State [29, 30] with [352 * 288] frame resolution, and an encoding rate of 30 frames per seconds (fps). This work also adds audio frames, which is 21.6 as in [13].

Two independent instances of the video conferencing application are used to stream the separate and distinct video and audio components of the Tokyo Olympics movie. These two applications configured to work simultaneously stream in the profile configuration [31]. The key parameters of this application configuration are the frame inter-arrival time and frame size. The incoming inter-arrival times are configured to the video and audio frame rates of 30 and 21.6, respectively. It should be noted that the outgoing inter-arrival time remains set to ''none'' in order to achieve unidirectional streaming from the video server. Furthermore, the frame size parameters are configured to explicitly script the video and audio traces.

# 6. RESULTS AND DISCUSSION

Sixty six scenarios were simulated, and the results are collected and summarized in three scenarios depending on different video codec of video application and varying path loss models, and for different types of service classes. For each scenario, the types of modulation and coding schemes are choosing one at a time to obtain one set of simulation results for the different performance measures of packet loss, packet delay, packet jitter, and traffic load throughput.

## 6.1 Scenario 1: Different video codec of video application

This subsection shows the simulation results of three scenarios under this category. Each scenario used different video codec under different modulation and coding scheme in each cell. A path-loss model is chosen as free space and service class as rtPS are considered, and kept constant. This simulation is used to evaluate the performance parameters, namely: packet jitter, packet E2E delay, data drop, and throughput of the mobile node.

The average packet jitter, and average E2E delay with various modulation and coding schemes are shown in Figures 4(a) – 4(b). Figure 4(a) shows the average variation of jitter for audio/video IPTV over Fixed WiMAX networks. For different coding, video quality is best if the jitter is zero. As shown in Figure 4(a), average audio/video jitter is approximately zero for higher modulation and coding scheme (MCS) (16 QAM, and 64 QAM) whereas QPSK has a worse average variation of jitter for Movie coded by AVC codec. From the results in Figure 4(a), it is observed that WiMAX using higher MCS (16 QAM, 64 QAM) as a modulation technique shows better jitter compared with other MCS (QPSK). It is also observed that video coded by SVC, and MPEG-4 has a better average jitter compared with the AVC codec. Therefore, video codec by SVC is the best for deploying IPTV. Average End-to-End delay for different video codec under MCS is shown in Figure 4(b), as it can be seen that the average E2E delay of different video codec gives lower packet E2E delay for audio/video IPTV when codec by SVC, and MPEG-4 that under all modulation and coding MCS.

(a)

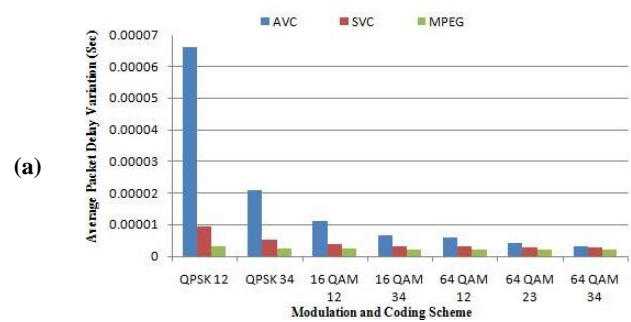

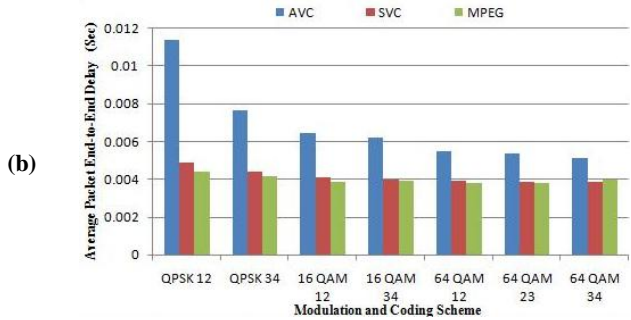

**(b)**

**Figure 4: (a) Average video jitter and**
**(b) Average packet End-to-End delay**

As shown in Figure 5(a), the average data drop is significantly higher when video codec by AVC codec. The effect of data drop naturally decreases the average WiMAX throughput as shown in Figure 5(b). From Figure 5(a), it is observed that the data dropped is very low for SVC video codec for all modulation and coding schemes. Whereas, the other different video codec (AVC, and MPEG-4) has more data dropped. Figure 5 (b), shows the average subscriber station (SS) WiMAX throughput. The average throughput for SVC is higher compared with its data dropped as shown in Figure 5 (a). Whereas, another codec has more throughput but also has more data dropped. According to the results as in Figures 5(a) -5 (b), it is observed that SVC codec is the best codec used to deploy IPTV over WiMAX, which has better performance (high throughput, low data dropped) under all modulation techniques compared with other video codec. In conclusion, transmitting SVC encoded videos over WiMAX networks is an effective solution for deploying IPTV.

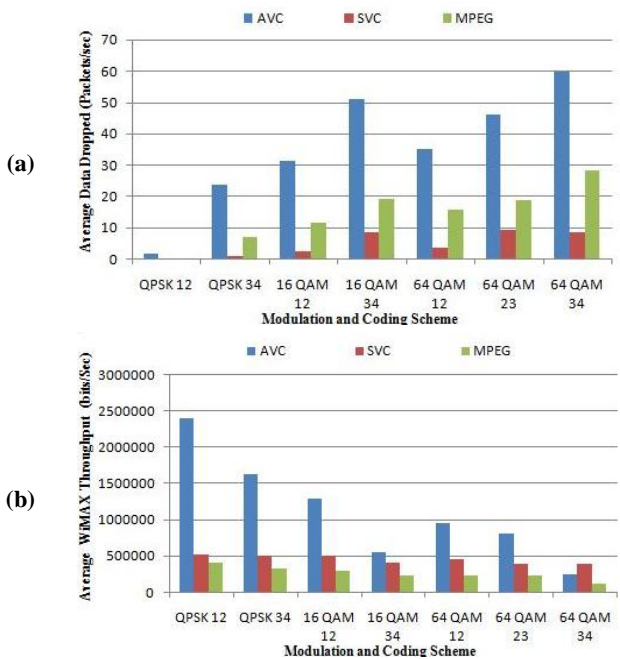

**Figure 5: (a) Average packet data dropped from SS node**
**(b) Average WiMAX throughput for SS node**

## 6.2 Scenario 2: Mobile node with different path loss

This subsection discusses the simulation results of twenty eight scenarios, performance parameters of each scenario observed for various modulation and coding schemes with respect to various path-loss models. It is considered in this category keeping the video codec with SVC codec and scheduling service classes as rtPS.

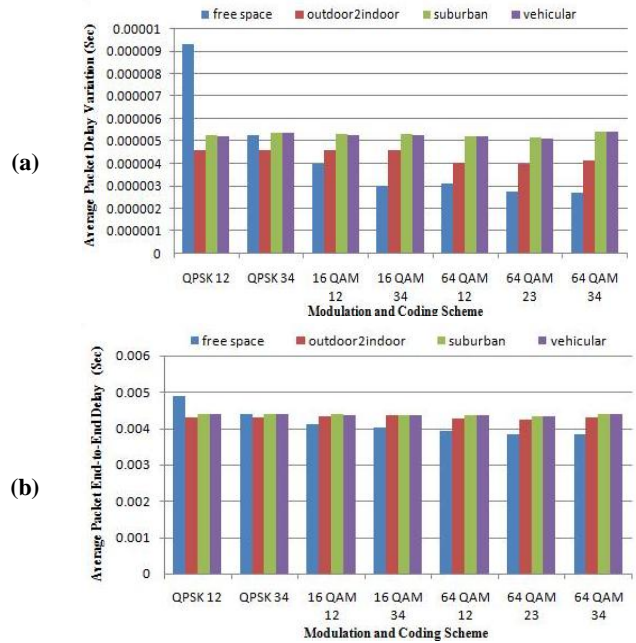

**(a)**

**(b)**

**Figure 6: (a) Average video jitter**
**(b) Average packet End-to-End delay**

In this case, fixed radius WiMAX networks are considered for all the path loss because, as in authors' knowledge, outdoor to indoor and pedestrian path-loss model is designed for small and micro cell WiMAX network. In the free space propagation model, fading and multi-path propagation phenomena wasn't considering in this work. Thus, path-loss would be very nominal and the received signal-to-noise ratio (SINR) would be ideal as can be seen from Figure 6 (a) -(b). It shows that free space path loss has less packet jitter and also, less E2E packet drop for all modulation and coding scheme except for the QPSK. Similarly, Figure 7 (b) shown the throughput for the free space propagation model is highest for all MCS. At the same time, it is considered that the suburban fixed model at the hilly terrain with high tree density that implies very high path-loss due to scattering and multi-path propagation of radio signals while for vehicular model, moderately flat terrain was considered so the path-loss would be less than that of the suburban fixed model. As vehicular model experiences very high packet drop compared with the others, it gives the lowest throughput compared with other propagation model except outdoor to indoor and pedestrian that can be observed from Figure 7 (b). Path-loss of free space is lowest; hence, the reduction of SINR with the distance from BS is less which leads to better throughput, lower packet jitter, lower packet E2E delay, and lower packet data dropped from all varies MCS as shown in Figures 6, and 7.

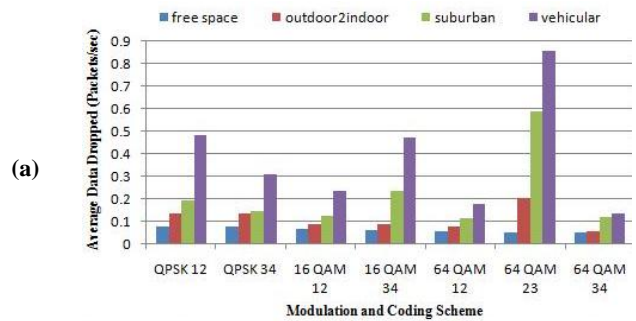

**(a)**

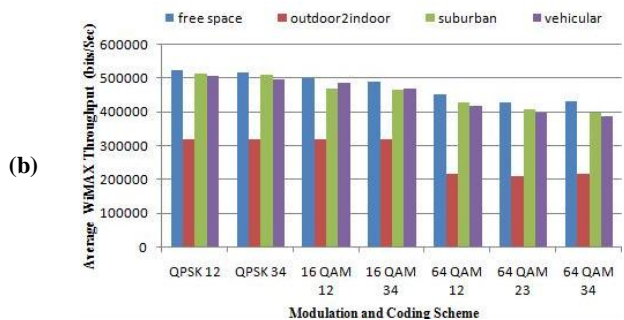

**(b)**

**Figure 7: (a) Average packet data dropped from SS node**
**(b) Average WiMAX throughput for SS node**

## 6.3 Scenario 3: Mobile node with different classes

This subsection exhibits the simulation results of 35 scenarios under this category where video codec and path loss kept constant. The video codec of SS is used SVC codec while path-loss model as free space is considered. Different service classes are used in this category under various modulations and coding scheme to obtain the performance metrics such as packet delay variation, packet End-to-End delay, data drooped for a SS mobile node, and WiMAX throughput for the SS mobile node.

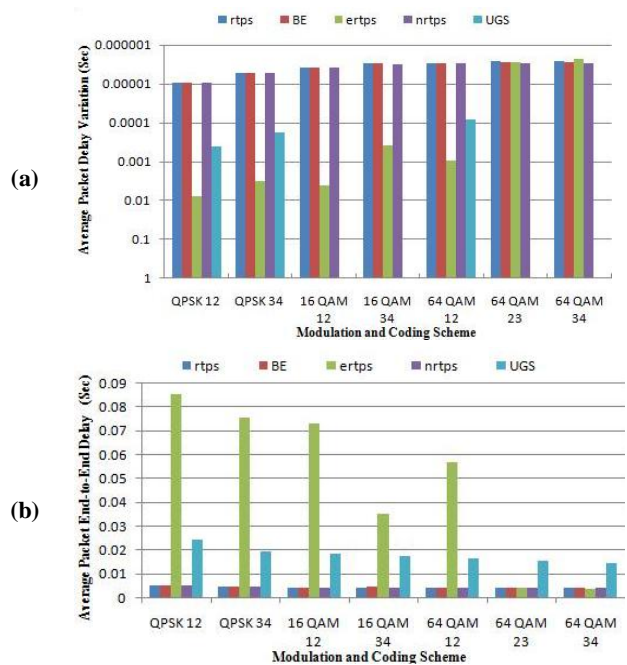

**(a)**

**(b)**

**Figure 8: (a) Average video jitter**
**(b) Average packet End-to-End delay**

It is known that UGS and ertPS were designed to support VoIP [32]. UGS is designed and used commonly for Constant Bit Rate (CBR) [33]. Figure 8 (a) -(b) shown that UGS and ertPS have more packet jitter delay and E2E delay. Similarly, Figure 9 (a) shows UGS, and ertPS have more packet drop for all modulations and coding schemes, that give less throughput as can be seen in Figure 9 (b). Figures 8 (a) -(b) show packet delay jitter and packet E2E delay for different service classes: as can be seen from these Figures rtPS, nrtPS, and BE have given the best performance which reveals that for all modulation and coding, all service classes give equal packet jitter, and equal packet E2E delay. Similarly, in Figure 9 (b), it is observed that rtPS classes providers have greater throughput than other classes nrtPS, and BE. Besides, it has lower packet dropped as can be seen in Figure 9 (a). The reason is that rtPS is designed for streaming Audio or Video.

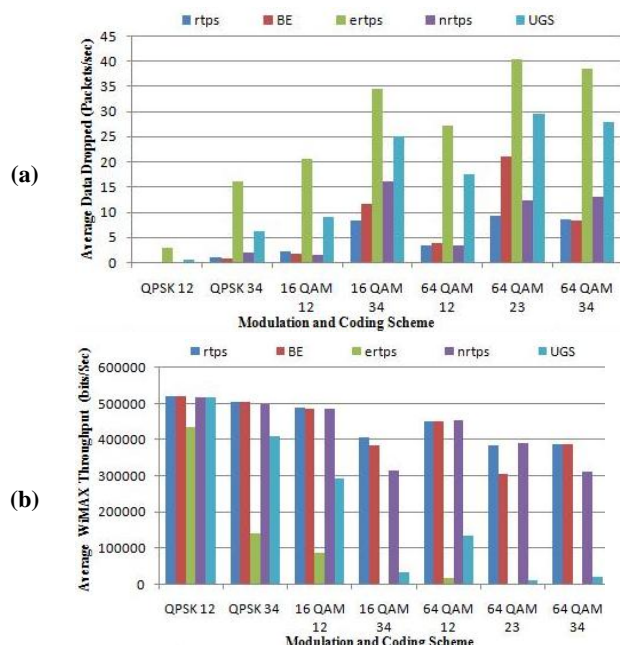

**(a)**

**(b)**

**Figure 9: (a) Average packet data dropped from SS node**
**(b) Average WiMAX throughput for SS nosde**

## 7. CONCLUSION

In this work, performance study of IPTV over Fixed WiMAX network considering different modulation and coding schemes have been presented under different key system parameters including video coding, path-loss models, and MAC service classes. The performance has been evaluated in terms of average packet jitter, average packet E2E delay, average throughput, and average data-dropped. OPNET simulation results show that SVC outperforms other video codec schemes. Also results show that the free space path loss is the best propagation model for deploying A/V video application over different fixed mobile node whereas vehicular model yields the poorest performance giving the highest packet drop rate. Moreover, simulation results show that rtPS scheduling service class is the most appropriate scheduling service for A/V video application. As a future study, study the impact of mobility on video quality, and also the impact of using multicast SVC multilayer adaptation scheme on enhancing the performance of video streaming over mobile WiMAX.